# How large is the error effect when summing or averaging nonlinear field normalization citation counts at the paper level?

Limi Tang[a,1]

[a]School of Information, Shanxi University of Finance & Economics, 030006 Taiyuan, China

[1]Corresponding author. Email: ican1118@163.com

**Abstract:**
　　Summing or averaging nonlinearly field-normalized citation counts is a common but methodologically problematic practice, as it violates mathematical principles. The issue originates from the nonlinear transformation, which disrupts the equal-interval property of the data. Such unequal data do not satisfy the necessary conditions for summation. In our study, we normalized citation counts of papers from all sample universities using six linear and nonlinear methods, and then computed the total and average scores for each university under each method. By benchmarking against raw citations and linear normalized scores, we explore how large the error effect is from summing or averaging the nonlinear field normalized citation counts. Our empirical results indicate that the error exists but is relatively small. We further found that the magnitude of the error is significantly influenced by whether the sample publications are homogeneous or heterogeneous. This study has significant implications for whether the results obtained through nonlinear methods on a single level can be directly summed or averaged when calculating the overall impact of a research unit.

# 1 Introduction

In the academic community, the measurement of scholarly influence is one of the key indicators for assessing the quality of research output. Citation count of publication is a common metric for evaluating the scholarly impact and research contributions of countries, institutions, and individuals. However, due to citation practices varying across fields, direct comparisons of raw citation counts are impractical (Bornmann et al., 2008). Field normalization methods address this issue by applying mathematical transformations to cross-field citations, enabling more accurate comparisons of scholarly impact (Waltman, 2016).

Various field normalization methods have been proposed in the scientometrics field. Zhang et al. (2015) proposed that these methods could be classified into linear and nonlinear methods based on whether the transformation is linear. Wang (2024) provided a more detailed definition of linear and nonlinear field normalization methods and categorized common field normalization methods. He identified the median-based method, z-score method, and mean-based method, et al., as linear field normalization methods, while methods including the percentile rank method, logarithmic z-score method, and Normalized Logarithmic Citation Score (NLCS) method, et al., are classified as nonlinear field normalization methods. He also noted that some methods, such as the citing-side normalization method and exchange rate method, are more difficult to categorize.

Zhang et al. (2015) also proposed that linear normalization methods, through linear transformations, ensure that citation counts satisfy the equal interval measurement requirement, making them suitable for summing or averaging operations. In contrast, nonlinear normalization methods alter the equal interval property of citation counts, and performing arithmetic operations directly on these data may lead to distorted results. Other scholars have also pointed out the drawbacks of nonlinear normalization methods, such as the Percentile rank method (Donner, 2022; Zhou and Zhong, 2012; D'Agostino et al., 2017). Similar issues have been noted in other fields, such as computer science, ergonomics, and operations management (Anjum and Perros, 2011; Kreifeldt and Nah, 1995; Van Hecke, 2010; Albin, 2017, 2019). Nevertheless, the misuse of nonlinear field normalized indicators remains prevalent in academic research practices (Lundberg, 2007; Maffahi and Thelwall, 2021; Bornmann, 2020; Thelwall and Maffahi, 2020; Caldwell et al., 2024).

Wang (2024) rigorously proved through mathematical theorems that linear transformations and equal intervals are both necessary and sufficient conditions for each other. He noted that summing or averaging nonlinear normalized citation counts with unequal intervals is mathematically unjustified, and can produce misleading results. Through empirical research, he confirmed that such errors exist and warned that they could affect the accuracy of university research impact rankings. He thus concluded that nonlinear field normalized citation counts should not be summed or averaged. In the latest study, he employed mathematical formulas to formally prove the theoretical proposition that nonlinearly normalized citation data cannot be summed. (Wang and Zhang, 2025)

However, Haunschild and Bornmann (2024) have offered a different perspective on Wang's conclusion. They argued that Wang's study lacks empirical support and does not correlate computational results with external criteria, such as peer review outcomes. In response, Wang (2025) clarified that his study aims to mathematically demonstrate why normalized citation counts cannot be summed, not to conduct empirical research. He emphasized that the principle preventing the aggregation of normalized citation data is independent of external validation. Moreover, using peer review to evaluate whether nonlinear citation counts can be summed is impractical, since it is not designed for such validations.

A critical issue remains: summing or averaging citation counts that lack the mathematical basis for such operations (i.e., non-equidistant data) is fundamentally flawed. It is like adding temperatures in Celsius and Fahrenheit directly. They both measure temperature, but use different scales and zero points, making such an addition meaningless. We believe understanding the mathematical principles behind normalization is essential. However, there is still a lack of systematic empirical research on the extent of errors introduced when such summing or averaging is applied to paper-level citation data.

In this study, we aim to systematically investigate the effects of the errors introduced by performing summing or averaging on nonlinear normalized scores through large-scale real citation counts. The specific experimental designs are as follows:

**1.Single-field experiment**: In this part, we investigate the errors resulting from summing or averaging the normalized citation counts of individual papers within a single discipline, compared to raw counts. Limiting the scope to a single field eliminates the confounding effects of cross-disciplinary differences, allowing for a more precise assessment of the impact of nonlinear normalization methods and a clearer evaluation of their mathematical properties.

**2.Cross-field experiment**: In this experiment, we focus on comparing the errors the linear and nonlinear field normalized scores after summing or averaging of individual papers across all fields from various universities. This approach better reflects real academic evaluation, given that university publications are inherently cross-disciplinary.

Through these experiments, we directly address a critical and practical question: how large is the error effect from summing or averaging nonlinear normalized scores? The results provide key guidance for applying nonlinear normalization methods in academic evaluations, especially at the institutional level.

## 2 Data and Methods

### 2.1 Data Source

This study utilizes bibliometric data from the InCites database, classified by the Essential Science Indicators (ESI) scheme to ensure comparability and avoid duplicate records. We restricted the institution type to "academic," the document type to "article," and the publication year to 2014. The data were collected between December 20, 2024, and January 9, 2025, ensuring a 10-year citation window for all the sample paper. The data is divided into three parts as follows.

**(1) Data for all papers across 22 fields.** We obtained the raw citation counts for all articles published in 2014 across the 22 ESI fields. For each field, we calculated the basic values of citation counts, such as the average, median, and standard deviation. For fields with over 50,000 documents, we obtained the complete dataset by applying filters, such as country or journal of publication.

**(2) Single-field experiment data.** We restricted the research area to four fields: Physics, Chemistry, Engineering, and Economics & Business. These were chosen for their representation of natural sciences, engineering, and social sciences, as well as their well-established systems, high research output, and academic influence. Their data stability ensures the broad applicability of our findings. From each field, we selected the top 1,000 universities by publication output to collect citation data.

**(3) Cross-field experiment data.** This dataset is fundamentally distinguished from the single-field experiment by its lack of disciplinary constraints. Based on an extension of the previous criteria, we removed field restrictions and directly selected the top 1,000 universities by total publication output across all disciplines, from which we collected the citation data.

**2.2 Normalization Methods and Statistical Analysis**

To thoroughly analyze the errors introduced by summing or averaging nonlinear field normalized citation counts, we used six different field normalization methods. These include three linear normalization methods: the Mean-based method (Radicchi et al., 2008; Waltman et al., 2011), the Median-based method (Leydesdorff & Opthof, 2011), and the Z-score method (Zhang et al., 2014) (see Table 1). The other three are nonlinear methods: the Percentile rank method (Bornmann,2013; Bornmann et al.,2013; Zhang et al., 2015), the Normalized Log Citation Score (NLCS) method (Thelwall, 2017), and the Logarithmic z-score method (Lundberg, 2007) (see in Table 2). These methods represent some of the most established and authoritative approaches in academic evaluation and can be directly implemented using existing citation databases.

**Table 1** Three linear field normalization methods used in this study.

| Name of method | Calculation Formula |
| --- | --- |
| Mean-based method | $Y=\dfrac{X}{m}$, $m$ is the average citation count of papers with the same field as the given paper. |
| Median-based method | $Y=\dfrac{X}{M}$, $M$ is the median citation count of papers with the same field as the given paper. |
| Z score method | $Y=\dfrac{(X-m)}{sd}$, $m$ is defined as above; $sd$ is the standard deviation of citation counts of papers with the same field as the given paper. |

*Note:* $Y$ represents papers' normalized scores by using each normalization method; $X$ is the raw citation counts of papers.

**Table 2** Three nonlinear field normalization methods used in this study.

| Name of method | Calculation Formula |
|---|---|
| Percentile rank method | $Y=\frac{n}{N} \cdot 100\%$, <br> $n$ is number of papers with citation counts ≤ the given paper (calculated using the Countif function). $N$ is the total number of papers with the same field as the given paper. |
| Normalized Log Citation Score (NLCS) method | $Y=\frac{ln(X+1)}{m_{[ln]}}$, <br> $m_{[ln]}$ is the mean of the log-transformed citation counts (plus one) distribution. |
| Logarithmic z-score method | $Y=\frac{ln(X+1)-m_{[ln]}}{Sd_{[ln]}}$, <br> $m_{[ln]}$ is defined as above. $Sd_{[ln]}$ the standard deviation of the log-transformed citation counts (plus one) distribution. |

*Note:* The definitions of Y and X are as described in Table1.

When standardizing the data, we first calculated the basic values of 22 fields, as mentioned above. Then, we use the formulas in Table 1 and 2 to calculate six types of normalized citation counts of each paper from the sample universities in both single and cross fields contexts. This process refers to apply six normalization methods to standardize the citation counts of each paper from each sample university. After that, we sum and average the normalized citation counts of per paper to obtain the overall normalized citation scores for each university. These scores reflect the overall impact of the university's research output (i.e., Aggregate Performance, AP). The formulas for the two operations are as follows,

a) Summing the normalized citation counts for each paper of the sample universities,

$$AP_1 = \sum_{i=1}^{n} C_i, \quad (1)$$

b) Averaging the normalized citation counts for each paper of the sample universities,

$$AP_2 = \frac{1}{n}\sum_{i=1}^{n} C_i, \quad (2)$$

where $n$ is the total number of papers published by a university and $C_i$ denotes the normalized citation counts for the *i-th* paper of a university.

The final phase of analysis involved using scatter plots to visually examine the distribution of universities' normalized scores and their rankings, in addition to calculating their Pearson and Spearman correlation coefficients. Subsequently, we assessed the changes in ranking that occurred after the summation or averaging of normalized citation counts. The single-field experiment used raw citation counts as a reference, conversely, the cross-field experiment employed the indictors of three linear normalization methods for this purpose.

## 3 Results and analysis
### 3.1 Single-field normalization experiment

Field normalization enables cross-field impact comparison, but methods differ mathematically: linear ones preserve additivity, while nonlinear ones distort equal-interval properties and lose it (Wang, 2024). To validate this theory, we begin with our study within four single fields. This approach, combined with a homogeneous sample (academic institutions, articles, 2014), removes the confound of field heterogeneity, thus allowing a purer assessment of the methods' intrinsic properties.

**(1) Pearson and Spearman correlation coefficient of indicators in four fields**

Tables 3 and 4 present the Pearson correlation between normalized scores (via $AP_1/AP_2$) and raw citations, and the Spearman correlation between their rankings. In all four single fields, linear normalization shows perfect correlation (1.000, $p = 0.000$) with raw citations. Nonlinear methods show weaker correlations, especially under averaging, the lowest being 0.769 for the Percentile rank score in Chemistry. Notably, the correlations for NLCS and Log z-score with raw citations are the same, regardless of the aggregation method. This is due to their similar computational logic (see Table 2).

**Table 3**
Pearson correlation coefficients between **summarized** field normalized scores and raw citation counts，and Spearman correlation coefficients between the ranking for these indicators for all sample universities.

| Field | Metric Type | Mean-based | Median-based | Z-score | Percentile Rank | NLCS | Log z-score |
|---|---|---|---|---|---|---|---|
| Physics | Score | 1.000** | 1.000** | 1.000** | .962** | .962** | .962** |
| | Ranking | 1.000** | 1.000** | 1.000** | .962** | .962** | .962** |
| Chemistry | Score | 1.000** | 1.000** | 1.000** | .962** | .957** | .957** |
| | Ranking | 1.000** | 1.000** | 1.000** | .964** | .957** | .957** |
| Engineering | Score | 1.000** | 1.000** | 1.000** | .978** | .976** | .976** |
| | Ranking | 1.000** | 1.000** | 1.000** | .959** | .957** | .957** |
| Economics & Business | Score | 1.000** | 1.000** | 1.000** | .941** | .943** | .943** |
| | Ranking | 1.000** | 1.000** | 1.000** | .949** | .948** | .948** |

Note: **. Significant at the 1% significance level. All the *P-value<0.001*.

**Table 4**
Pearson correlation coefficients between **averaged** field normalized scores and raw citation counts, and Spearman correlation coefficients between the ranking for these indicators for all sample universities.

| Field | Metric Type | Mean-based | Median-based | Z-score | Percentile Rank | NLCS | Log z-score |
|---|---|---|---|---|---|---|---|
| Physics | Score | 1.000** | 1.000** | 1.000** | .911** | .934** | .934** |
| | Ranking | 1.000** | 1.000** | 1.000** | .936** | .953** | .953** |
| Chemistry | Score | 1.000** | 1.000** | 1.000** | .769** | .792** | .792** |
| | Ranking | 1.000** | 1.000** | 1.000** | .889** | .904** | .904** |
| Engineering | Score | 1.000** | 1.000** | 1.000** | .795** | .827** | .827** |
| | Ranking | 1.000** | 1.000** | 1.000** | .864** | .888** | .888** |
| Economics & Business | Score | 1.000** | 1.000** | 1.000** | .791** | .831** | .831** |
| | Ranking | 1.000** | 1.000** | 1.000** | .865** | .896** | .896** |

Note: **. Significant at the 1% significance level. All the *P-value<0.001*.

Clearly, within a single field, linearly normalized scores and rankings remain perfectly consistent with raw citation counts upon aggregation (summing or averaging). Nonlinear indicators, however, show marked deviations in both scores and rankings. This error is not influenced by factors such as publication year, field heterogeneity, or document type but originates from the nature of nonlinear normalization method itself. It strips the raw data of its equal-interval property, destroying additivity. However, when calculating the sum or average of university's paper normalized scores, the core operations inevitably involve addition of individual paper citation counts. It can be concluded that the direct cause of the observed bias in our experiment is the forced summation of nonlinear field normalized citation counts, which is inherently non-additive.

### (2) Scatter plots of indicators in four fields

To streamline the presentation, the Mean-based score and the Percentile rank score are selected as representative indicators of the linear and nonlinear normalization methods, respectively. The subsequent scatter plots depict the correlation of these scores and their rankings with those of the raw citation counts across the four fields. Additionally, scatter plots of the other four normalization methods are available in Appendix A, Fig.8- Fig.11.

Figs. 1 and 2 reveal a distinct contrast. Across all fields, Mean-based scores and their rankings consistently form a tight linear alignment with raw citation counts (see subfigures a-d1, a-d3), thereby demonstrating high consistency. Conversely, Percentile rank scores and their rankings exhibit substantial dispersion (see subfigures a-d2, a-d4), a pattern that is markedly pronounced under averaging. The scatter plots thus provide visual confirmation for the correlation coefficients in Tables 3 and 4, demonstrating the discrepancy between the Percentile rank indicator and the raw data. Notably, the magnitude of this error is slightly smaller when normalized scores are aggregated by summation compared to averaging.

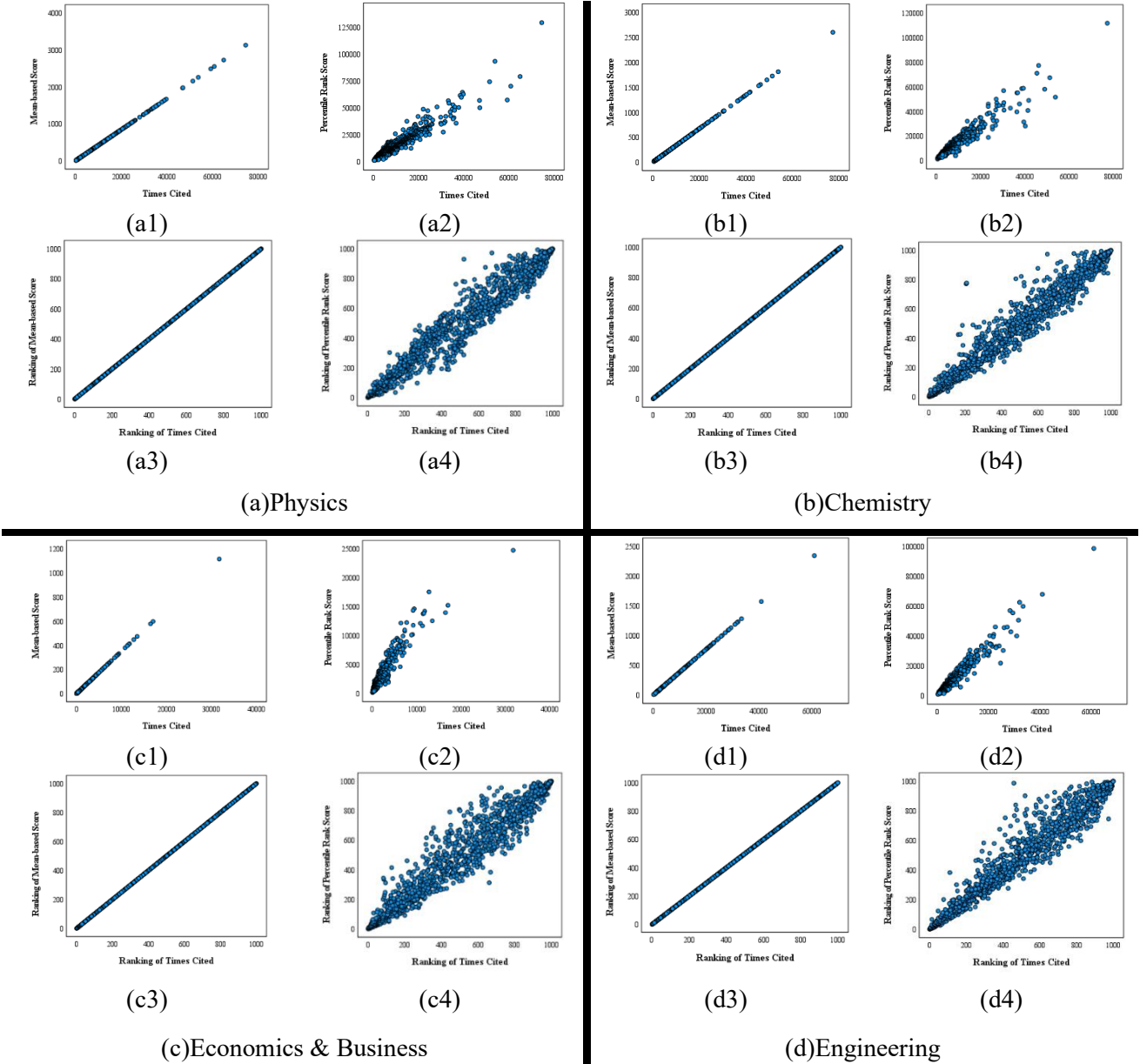

Fig.1. Correlation between the **summarized Mean-based and Percentile rank score** against raw citation counts and correlation between the rankings of these indicators for sample universities in different fields.

Note: 1. (a1)- (a2), (b1)- (b2), (c1)- (c2), (d1)- (d2) represent of the correlation of the Mean-based and Percentile rank *score against raw citation counts* for sample universities in Physics, Chemistry, Economics & Business and Engineering, respectively;

2. (a3)- (a4), (b3)- (b4), (c3)- (c4), (d3)- (d4) represent the correlation of Mean-based and Percentile rank score *rankings versus raw citation counts' rankings* for sample universities in Physics, Chemistry, Economics & Business, and Engineering, respectively;

3. Times Cited in this paper means raw citation counts.

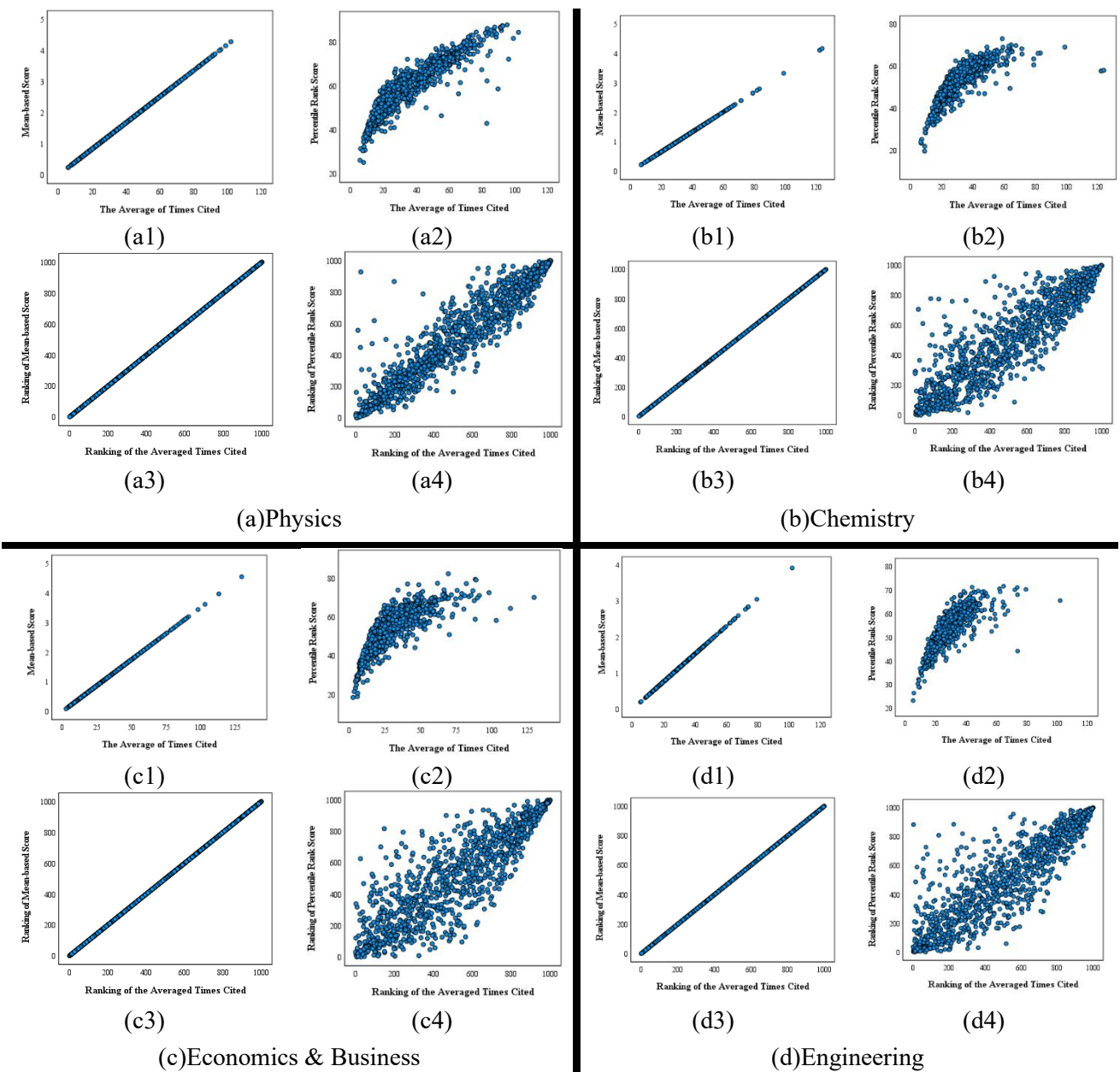

**Fig.2.** Correlation between the **averaged Mean-based and Percentile rank score** against the averaged raw citation counts and correlation between the rankings of these indicators for sample universities in different fields.

Note: (a1)- (d4) represent correlations consistent with Fig.1.

### (3) Ranking changes of indicators in Physics

To more intuitively illustrate the ranking distortions caused by summing or averaging nonlinearly normalized citation counts, we systematically analyzed ranking changes for all sample universities across the four fields. The ranking shifts for the top 20 most productive universities in Physics are shown here. Complete results for all universities are provided in the Excel attachment named "Single-field Data" in the *Supplementary Materials*.

Tables 5 and 6 present the results for the representative Mean-based score (with other linear indicators in Appendix Tables 12–13). They show that the rankings from linear normalization are nearly unchanged from the raw citation count rankings (see Rank C1 columns) after both summation and averaging. In contrast, nonlinear normalization methods cause substantial ranking fluctuations, especially under averaging (Table 5). For example, Lomonosov Moscow State University (No. 0009) jumps 159 positions using Percentile rank method, while No.0016 and No.0004 shift 153 and 148 positions respectively. Under the summing operation, the most pronounced change among top-20 institutions occurs at Huazhong University of Science & Technology (No. 0020), which declines 115 positions consistently across all three nonlinear methods. These results demonstrate that aggregating non-additive nonlinear data produces significant ranking distortions.

Additionally, the ranking changes under the NLCS and Logarithmic z-score methods (see Rank C3 and C4 columns) are nearly identical, consistent with our prior correlation analyses (Tables 3 and 4). These findings collectively demonstrate that linear normalization is reliable for aggregation, unlike nonlinear methods, which induce significant ranking distortions. This provides definitive empirical support for the mathematical principle that nonlinear normalized counts lack additivity and should not be summed.

**Table 5**
Ranking changes of the Top 20 universities with the most articles in Physics after **averaging** the normalization citation counts

| No. | TC (ave) | Rank0 | Mean score | Rank1 | Rank C1 | PR score | Rank2 | Rank C2 | NLCS score | Rank3 | Rank C3 | Log z score | Rank4 | Rank C4 |
|---|---|---|---|---|---|---|---|---|---|---|---|---|---|---|
| 0001 | 36.257 | 468 | 1.522 | 468 | 0 | 62.988 | 430 | -38 | 1.21 | 439 | -29 | 2.423 | 439 | -29 |
| 0002 | 35.555 | 480 | 1.493 | 480 | 0 | 61.802 | 465 | -15 | 1.191 | 470 | -10 | 2.385 | 470 | -10 |
| 0003 | 45.818 | 346 | 1.924 | 346 | 0 | 66.378 | 333 | -13 | 1.272 | 341 | -5 | 2.547 | 341 | -5 |
| 0004 | 34.414 | 502 | 1.445 | 502 | 0 | 55.327 | 650 | 148 | 1.087 | 625 | 123 | 2.178 | 625 | 123 |
| 0005 | 58.323 | 165 | 2.449 | 165 | 0 | 71.26 | 190 | 25 | 1.379 | 180 | 15 | 2.763 | 180 | 15 |
| 0006 | 35.453 | 482 | 1.488 | 482 | 0 | 58.198 | 572 | 90 | 1.13 | 560 | 78 | 2.263 | 560 | 78 |
| 0007 | 37.214 | 453 | 1.562 | 453 | 0 | 58.842 | 553 | 100 | 1.139 | 543 | 90 | 2.282 | 543 | 90 |
| 0008 | 60.491 | 143 | 2.54 | 143 | 0 | 69.984 | 228 | 85 | 1.356 | 218 | 75 | 2.716 | 218 | 75 |
| 0009 | 33.451 | 513 | 1.404 | 513 | 0 | 54.498 | 672 | 159 | 1.065 | 663 | 150 | 2.133 | 663 | 150 |
| 0010 | 35.658 | 478 | 1.497 | 478 | 0 | 61.146 | 486 | 8 | 1.182 | 486 | 8 | 2.367 | 486 | 8 |
| 0011 | 24.298 | 681 | 1.02 | 681 | 0 | 53.977 | 685 | 4 | 1.048 | 688 | 7 | 2.1 | 688 | 7 |
| 0012 | 22.478 | 723 | 0.944 | 723 | 0 | 47.969 | 842 | 119 | 0.954 | 828 | 105 | 1.911 | 828 | 105 |
| 0013 | 44.717 | 360 | 1.877 | 360 | 0 | 68.923 | 262 | -98 | 1.316 | 286 | -74 | 2.636 | 286 | -74 |
| 0014 | 32.663 | 523 | 1.371 | 523 | 0 | 59.264 | 544 | 21 | 1.145 | 533 | 10 | 2.294 | 533 | 10 |
| 0015 | 28.776 | 590 | 1.208 | 590 | 0 | 56.843 | 607 | 17 | 1.099 | 610 | 20 | 2.201 | 610 | 20 |
| 0016 | 73.225 | 49 | 3.074 | 49 | 0 | 71.04 | 202 | 153 | 1.385 | 175 | 126 | 2.773 | 175 | 126 |
| 0017 | 58.456 | 161 | 2.454 | 161 | 0 | 71.039 | 203 | 42 | 1.364 | 203 | 42 | 2.731 | 203 | 42 |
| 0018 | 45.272 | 355 | 1.901 | 355 | 0 | 67.984 | 287 | -68 | 1.304 | 305 | -50 | 2.611 | 305 | -50 |
| 0019 | 48.104 | 304 | 2.02 | 304 | 0 | 67.803 | 294 | -10 | 1.304 | 302 | -2 | 2.612 | 302 | -2 |
| 0020 | 19.751 | 776 | 0.829 | 776 | 0 | 50.363 | 781 | 5 | 0.98 | 793 | 17 | 1.962 | 793 | 17 |

Note: *No. 0001-0020* represent the top 20 universities with the highest publication outputs in Physics, with specific university names listed in Appendix Table 12; *Rank0-4* respectively refer to the rankings of average of raw citation counts (TC ave), Mean-based score, Percentile rank score, NLCS score and Log z-score; *Rank C1-4* respectively refer to the changes in rankings of Mean-based score, Percentile rank score, NLCS score, and Log z-score against the average of raw citation counts' rankings(Rank0).

**Table 6**
Ranking changes of the Top 20 universities with the most articles in Physics after **summing** the normalization citation counts

| No. | TC | Rank0 | Mean score | Rank1 | Rank C1 | PR score | Rank2 | Rank C2 | NLCS score | Rank3 | Rank C3 | Log z score | Rank4 | Rank C4 |
|---|---|---|---|---|---|---|---|---|---|---|---|---|---|---|
| 0001 | 74509 | 1 | 3128.14 | 1 | 0 | 129440.623 | 1 | 0 | 2486.27 | 1 | 0 | 4979.641 | 1 | 0 |
| 0002 | 53795 | 5 | 2258.496 | 5 | 0 | 93506.659 | 2 | -3 | 1801.46 | 2 | -3 | 3608.064 | 2 | -3 |
| 0003 | 51408 | 6 | 2158.282 | 6 | 0 | 74475.941 | 4 | -2 | 1427.052 | 4 | -2 | 2858.179 | 4 | -2 |
| 0004 | 38372 | 12 | 1610.986 | 12 | 0 | 61689.332 | 8 | -4 | 1212.307 | 8 | -4 | 2428.075 | 8 | -4 |
| 0005 | 64914 | 2 | 2725.309 | 2 | 0 | 79312.442 | 3 | 1 | 1535.14 | 3 | 1 | 3074.664 | 3 | 1 |
| 0006 | 39388 | 10 | 1653.642 | 10 | 0 | 64657.886 | 6 | -4 | 1255.222 | 6 | -4 | 2514.028 | 6 | -4 |
| 0007 | 39782 | 9 | 1670.183 | 9 | 0 | 62902.223 | 7 | -2 | 1217.968 | 7 | -2 | 2439.414 | 7 | -2 |
| 0008 | 60793 | 3 | 2552.296 | 3 | 0 | 70333.776 | 5 | 2 | 1362.816 | 5 | 2 | 2729.525 | 5 | 2 |
| 0009 | 33418 | 23 | 1403.001 | 23 | 0 | 54443.877 | 13 | -10 | 1064.113 | 13 | -10 | 2131.264 | 13 | -10 |
| 0010 | 33126 | 24 | 1390.742 | 24 | 0 | 56804.441 | 12 | -12 | 1097.674 | 11 | -13 | 2198.482 | 11 | -13 |
| 0011 | 22014 | 62 | 924.222 | 62 | 0 | 48903.039 | 19 | -43 | 949.852 | 19 | -43 | 1902.416 | 19 | -43 |
| 0012 | 19736 | 77 | 828.584 | 77 | 0 | 42116.871 | 26 | -51 | 837.794 | 24 | -53 | 1677.98 | 24 | -53 |
| 0013 | 38904 | 11 | 1633.322 | 11 | 0 | 59962.581 | 9 | -2 | 1144.865 | 9 | -2 | 2293 | 9 | -2 |
| 0014 | 28123 | 33 | 1180.699 | 33 | 0 | 51026.082 | 15 | -18 | 986.271 | 15 | -18 | 1975.359 | 15 | -18 |
| 0015 | 24028 | 48 | 1008.777 | 48 | 0 | 47463.843 | 22 | -26 | 917.425 | 22 | -26 | 1837.469 | 22 | -26 |
| 0016 | 59239 | 4 | 2487.054 | 4 | 0 | 57471.361 | 10 | 6 | 1120.08 | 10 | 6 | 2243.359 | 10 | 6 |
| 0017 | 46940 | 8 | 1970.7 | 8 | 0 | 57044.397 | 11 | 3 | 1094.978 | 12 | 4 | 2193.083 | 12 | 4 |
| 0018 | 35222 | 19 | 1478.739 | 19 | 0 | 52891.481 | 14 | -5 | 1014.211 | 14 | -5 | 2031.319 | 14 | -5 |
| 0019 | 36078 | 15 | 1514.677 | 15 | 0 | 50852.295 | 16 | 1 | 978.096 | 16 | 1 | 1958.986 | 16 | 1 |
| 0020 | 14774 | 150 | 620.263 | 150 | 0 | 37671.774 | 35 | -115 | 732.791 | 35 | -115 | 1467.675 | 35 | -115 |

Note: *No. 0001-0020, Rank0-4* and *Rank C1-4* as defined in Table 5.

While all previous analyses consistently verify the presence of an error between nonlinear field-normalized scores and raw citation counts, it is evident that the extent of this discrepancy remains relatively limited. The correlation coefficients for both scores and rankings stay above 0.7, and the scatter plots continue to exhibit an overall linear distribution with relatively high density, particularly under summation. It must be emphasized, however, that the inherent absence of additivity in nonlinear normalized data constitutes a fundamental flaw that cannot be disregarded.

### 3.2 Cross-field normalization experiment

The previous single-field experiment focused on the mathematical flaws of nonlinear field normalization methods. However, in practical research assessment, evaluation sets typically contain publications from multiple fields. Thus, in this section, we investigate the error effects introduced by nonlinear normalization methods in cross-field scenarios.

**(1) When the sample size is 1000 universities, we explore the errors between nonlinear and linear field normalization indictors.**

The correlations among indicators are summarized in Table 7 (for summation, AP1) and Table 8 (for averaging, AP2). In all following sections, "r" represents the Pearson correlation coefficient and "rho" the Spearman rank correlation coefficient. Here, we illustrate the interpretation of these metrics using the data from Table 7. The first 0.980 in the "r" column represents the Pearson correlation coefficient between the Mean-based score and Percentile rank score, while the 0.980 in the "rho" column represents the Spearman correlation coefficient between their rankings. The same interpretation applies to the following tables in this paper.

**Table 7**
Pearson correlation between linear and nonlinear field normalized scores **by $AP_1$** and Spearman correlation between the rankings of these indicators for all sample universities

| Linear | Nonlinear | Sample size | r | P-value(r) | rho | P-value(rho) |
|---|---|---|---|---|---|---|
| Mean-based score | Percentile Rank score | 1000 | .980** | 0.000 | .980** | 0.000 |
|  | NLCS | 1000 | .977** | 0.000 | .976** | 0.000 |
|  | Log z-score | 1000 | .974** | 0.000 | .971** | 0.000 |
| Median-based score | Percentile Rank score | 1000 | .980** | 0.000 | .979** | 0.000 |
|  | NLCS | 1000 | .977** | 0.000 | .975** | 0.000 |
|  | Log z-score | 1000 | .973** | 0.000 | .970** | 0.000 |
| Z-score | Percentile Rank score | 1000 | .994** | 0.000 | .994** | 0.000 |
|  | NLCS | 1000 | .993** | 0.000 | .992** | 0.000 |
|  | Log z-score | 1000 | .991** | 0.000 | .989** | 0.000 |

Note: Linear and Nonlinear mean the linear field normalized indicators and the nonlinear field normalized indicators, respectively.
**. Significant at the 1% significance level.

**Table 8**
Pearson correlation between linear and nonlinear field normalized scores **by $AP_2$** and Spearman correlation between the rankings of these indicators for all sample universities

| Linear | Nonlinear | Sample size | r | P-value(r) | rho | P-value(rho) |
|---|---|---|---|---|---|---|
| Mean-based score | Percentile Rank score | 1000 | .918** | 0.000 | .944** | 0.000 |
| | NLCS | 1000 | .928** | 0.000 | .951** | 0.000 |
| | Log z-score | 1000 | .931** | 0.000 | .953** | 0.000 |
| Median-based score | Percentile Rank score | 1000 | .911** | 0.000 | .937** | 0.000 |
| | NLCS | 1000 | .925** | 0.000 | .947** | 0.000 |
| | Log z-score | 1000 | .924** | 0.000 | .946** | 0.000 |
| Z-score | Percentile Rank score | 1000 | .932** | 0.000 | .951** | 0.000 |
| | NLCS | 1000 | .941** | 0.000 | .957** | 0.000 |
| | Log z-score | 1000 | .943** | 0.000 | .959** | 0.000 |

Note: Linear and Nonlinear as defined in Table 7.
**. Significant at the 1% significance level.

Overall, we can see that whether by $AP_1$ or $AP_2$, the correlation coefficients between nonlinear and linear normalized scores and their rankings are all greater than 0.9, with even stronger correlations observed in the summation case. Notably, under summation, the correlations between the z-score (i.e., a typical representative of linear field normalization methods) and the nonlinear indicators generally above 0.99, with one value at 0.989. Although averaging reduces these correlations, it is important to note that the z-score still exhibits higher correlations with nonlinear methods than the other linear methods do.

The high correlation between Z-score and nonlinear normalized scores can be explained not only by the inherent limitations of the Z-score method but also by the nature of summation operations. According to Zhang et al. (2014), the Z-score method is suitable for normal distributions but inefficient under skewed distributions. Since real citation data, including our sample, display pronounced right-skewness (Seglen, 1992), the Z-score's efficacy is reduced. Additionally, in summation, the dominant effect of high-cited papers diminishes the contribution of less-cited ones, thereby elevating the correlation between z-score and nonlinear indicators in summation relative to averaging.

Comparing Tables 7 and 8 shows that correlations between nonlinear and linear methods differ between summation and averaging. By summing, the correlation between the scores and rankings is highly consistent; while by averaging, rank correlations (rho) are generally higher than score correlations (r). This suggests that, although averaging increases the scores' error, the university ranking results remain relatively stable.

As shown in Fig. 3(a)–(f), which display scatter plots between Mean-based and nonlinear normalized scores under $AP_1$ and $AP_2$, and Fig. 3(g)–(l), which present the corresponding ranking distributions, all subplots exhibit tightly clustered data. Scatter plots based on summation are denser than those based on averaging. In addition, the ranking distributions under summation (Fig. 3(g)–(i)) are more closely aligned with the diagonal, whereas those under averaging (Figs. 3(j)–(l)) exhibit greater dispersion. This pattern confirms that averaging introduces larger errors between nonlinear and linear indicators than summation does.

    To further validate our findings, we performed a comparative analysis using scatter plots (see Appendix B, Fig. 12 and 13) for the other two linear normalization indicators (i.e., Median-based score and Z-score) against the nonlinear indicators and their rankings. The resulting correlations closely match those obtained with the Mean-based method.

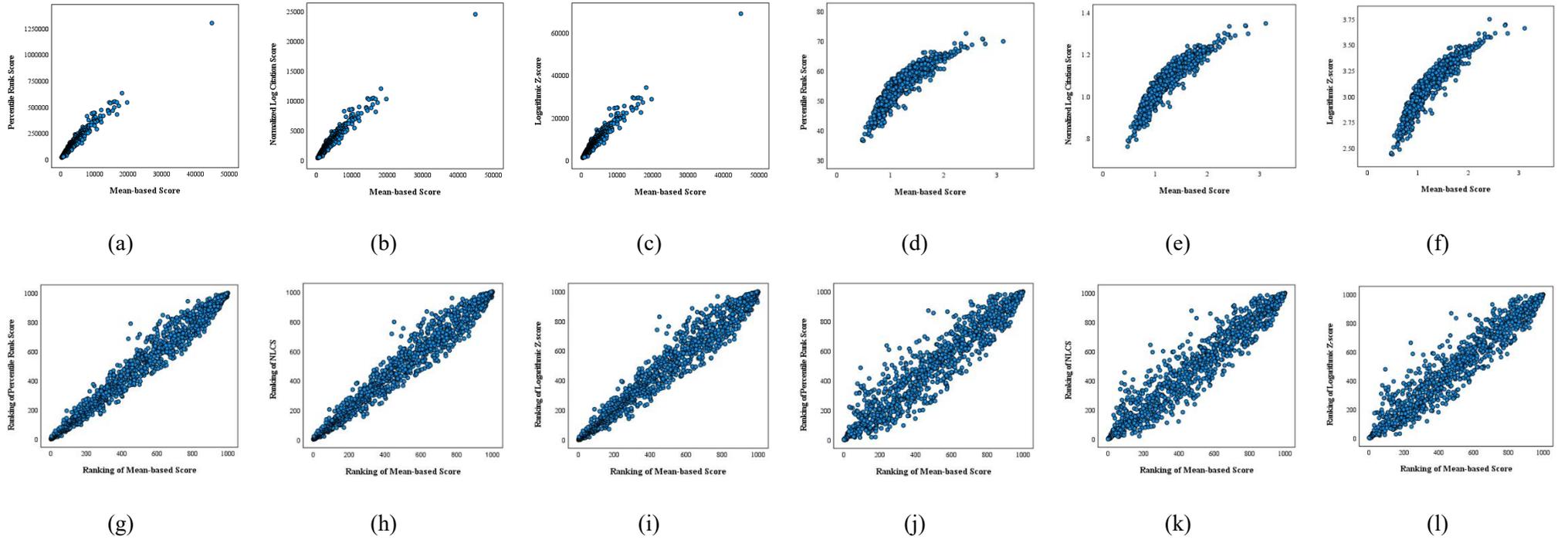

Fig.3. Correlation of **Mean-based score** against nonlinear field normalization score and the correlation of their rankings for all sample universities

Note: 1. (a)-(c) show the correlation between the Mean-based score and nonlinear field normalized *scores by summing* ($AP_1$), (d)-(f) show the correlation of these *scores by averaging* ($AP_2$).

2. (g)-(i) show the *ranking distribution* between the Mean-based score and nonlinear scores *by summing* ($AP_1$), (j)-(l) show the r*anking distribution* of these scores *by averaging* ($AP_2$).

**(2) When the sample size is 100, we explore the error between nonlinear and linear field normalization indictors.**

To investigate whether the high correlations and tightly clustered distributions are sample-size dependent, we extracted the top 100 universities from the existing cross-field data and analyzed the correlations of their normalized indicators. With a sample of 100 universities (Table 9), the Z-score shows lower correlation under summation but higher correlation under averaging compared to the 1000-university sample (Tables 7 and 8). Despite this, all correlations remain high with the lowest r and rho being 0.962 and 0.954, respectively. This indicates that a strong correlation persists between nonlinear and linear normalization even with a limited sample of top universities. This finding holds when the Median-based and Mean-based scores are used as benchmarks (see Appendix B, Tables 14 and 15).

**Table 9**
Pearson correlations between **Z-score** and nonlinear field normalized scores by $AP_1$ or $AP_2$ and Spearman correlation coefficient between the ranking for these indicators

| Operation | Nonlinear | Sample size | r | P-value(r) | rho | P-value(rho) |
|---|---|---|---|---|---|---|
| $AP_1$ | Percentile Rank score | 100 | .986** | <0.001 | .977** | <0.001 |
|  | NLCS | 100 | .983** | <0.001 | .967** | <0.001 |
|  | Log z-score | 100 | .977** | <0.001 | .954** | <0.001 |
| $AP_2$ | Percentile Rank score | 100 | .962** | <0.001 | .969** | <0.001 |
|  | NLCS | 100 | .970** | <0.001 | .975** | <0.001 |
|  | Log z-score | 100 | .971** | <0.001 | .974** | <0.001 |

*Note:* $AP_1$ and $AP_2$ as described in section 2.2; Nonlinear means the nonlinear field-normalized indicators.
**. Significant at the 1% significance level.

Figures 4 and 5 show similar scatter plot distributions for the 100- and 1000-university samples (Fig. 3). The main difference is that point density correspondingly declines with reduced sample size. In summary, our analysis indicates that although the correlation between nonlinear and linear normalized indicators changes when the sample is reduced to 100 universities, the overall difference is minor, and correlation coefficients remain high. This suggests that while the sample size affects the specific correlation values, it is not the key factor determining the strength of the correlation between linear and nonlinear normalized indicators.

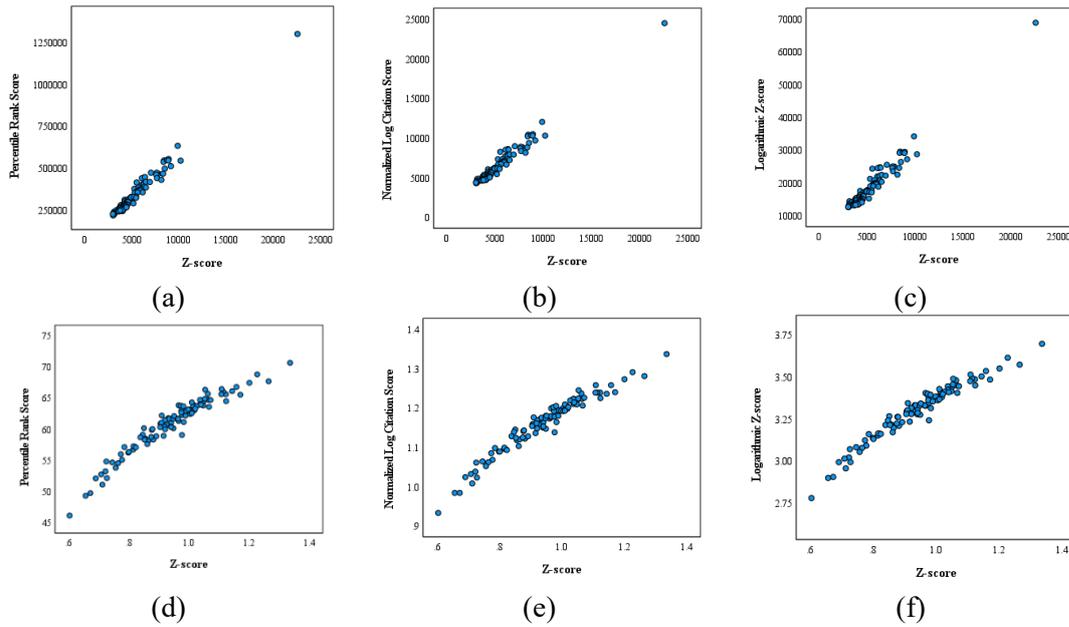

Fig. 4. Correlation of Z-score against nonlinear field normalized scores for 100 universities.
*Note*: (a)-(c) and (d)-(f) show the correlation of the Z-score against nonlinear field normalized scores by $AP_1$ and $AP_2$, respectively.

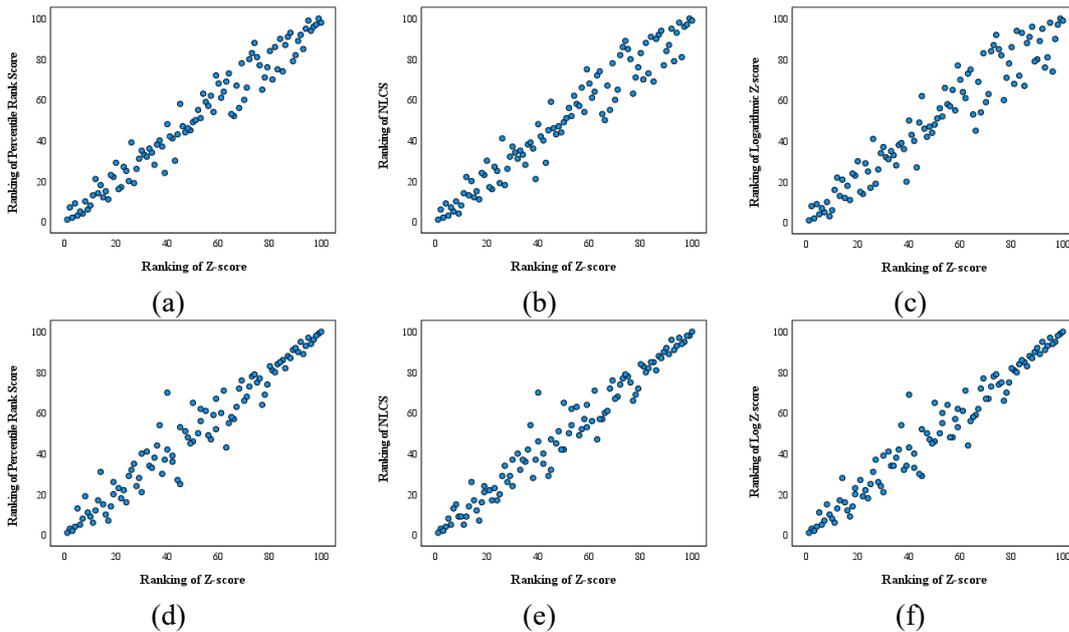

Fig.5. Correlation of the rankings of Z score and nonlinear field normalized scores for 100 universities.
*Note:* (a)-(c) and (d)-(f) show the ranking distribution among the Z-score and nonlinear field normalized scores by $AP_1$ and $AP_2$, respectively.

We also analyzed changes in the sample universities' rankings after summing and averaging their field-normalized citation counts. Due to length limitations, these results can be found in the "Cross-filed Data" Excel file in the *Supplementary Materials*.

The cross-field analysis indicates that the correlation between the linear and nonlinear field-normalized scores (via summation or averaging) is significant, with most correlation coefficients above 0.885. Despite this strong association, measurable errors are observed between the two types of indicators. The degree of error is smaller in cross-field compared to single-field experiment. A plausible explanation for this phenomenon is the adoption of linear normalized indicators as the benchmark in the cross-field experiment. As we discussed earlier, linear normalization method has limitations when handling skewed citation data. These limitations lead to a stronger correlation with the nonlinear normalized indicators.

In summary, our analysis across single-field and cross-field experiments indicates that the correlation between nonlinear and linear field-normalized scores remains high (>0.7) following summation or averaging, suggesting generally minor errors between the two. Nonetheless, these errors are consistent and non-zero. This finding provides direct empirical support for the theoretical framework established by Wang (2024) who demonstrated mathematically that nonlinear field normalization methods disturb the original equal intervals of citation counts, thereby rendering the normalization citation data non-additive. As posited, the forced summation or averaging of such data with unequal intervals is prone to introducing errors.

In practical research evaluation, it is crucial to fully recognize the inevitability of errors caused by methodological flaws, especially when conducting high-stakes assessments such as institutional rankings or resource allocation. We therefore argue that evaluators should prioritize methods which preserve data additivity when calculating overall impact. Methods that produce non-equivalent normalized data are prone to bias, and their results should be treated with caution to avoid skewed and impactful decisions.

## 4 Discussion

This study's experimental results demonstrate a strong correlation (>0.7) between nonlinear and linear field-normalized scores and their rankings. This finding is notably inconsistent with Wang's (2024) results, where the correlation between CNCI (i.e., Category Normalized Citation Impact) and AP (i.e., Average Percentile) scores and rankings reached only 0.556 at best. To analyze this difference, we sought to identify its underlying cause. Our further investigation revealed that whether document types are filtered during data collection is a critical factor influencing the correlation between these normalized scores. This point is also noted in the recent work of Shen (2025). The supporting experimental design and results are presented below.

In this section, we investigate how four factors (i.e., document type, publication year, filed classification schema, and sample size) affect the correlation and ranking stability between the linear (CNCI) and nonlinear (AP) field-normalized indicators, using a controlled-variable approach. We defined our experimental conditions first. We restricted the institution type to "academic" and combined two different time frames (2014 alone and 2014-2020), two field classification schemes (ESI and WoS), and two document type conditions ("articles" only and all types). This resulted in eight distinct data collection scenarios. For each scenario, we independently obtained the citation data of the top 2,000 institutions ranked by their number of Web of Science publications under those specific conditions. The data were organized into a single-year group and a multi-year group, with each containing the respective ESI and WoS sets which were further divided into filtered and unfiltered document conditions. All data were retrieved between March 31 and April 2, 2025. For each subset, we ranked the CNCI and AP scores of the top 100, top 1,000, and top 2,000 universities and calculated the consequent ranking shifts between the two metrics. The data can be found in the "Data of discussion section" folder of the *Supplementary Materials*,

Table 10
Correlation coefficients between CNCI and AP scores and their rankings for given sample universities within single-year group (in 2014)

| Schema | Document type | Sample size | r | p-value(r) | rho | p-value(rho) |
|---|---|---|---|---|---|---|
| ESI | Type-Filtered | Top100 | 0.949** | <0.001 | 0.963** | <0.001 |
| | | Top1000 | 0.919** | 0.000 | 0.943** | 0.000 |
| | | Top2000 | 0.878** | 0.000 | 0.916** | 0.000 |
| | Type-Unfiltered | Top100 | 0.559** | <0.001 | 0.465** | <0.001 |
| | | Top1000 | 0.650** | <0.001 | 0.663** | <0.001 |
| | | Top2000 | 0.693** | <0.001 | 0.739** | 0.000 |
| WoS | Type-Filtered | Top100 | 0.936** | <0.001 | 0.949** | <0.001 |
| | | Top1000 | 0.908** | 0.000 | 0.946** | 0.000 |
| | | Top2000 | 0.868** | 0.000 | 0.920** | 0.000 |
| | Type-Unfiltered | Top100 | 0.618** | <0.001 | 0.551** | <0.001 |
| | | Top1000 | 0.714** | <0.001 | 0.734** | <0.001 |
| | | Top2000 | 0.735** | 0.000 | 0.779** | 0.000 |

Note: The publication date for papers is set to 2014.
Top 100-2000 refer to the top 100 universities, top 1000 universities, and top 2,000 universities with the highest number of Web of science publications.
**, Significant at the 0.01 significance level.

Table 11
Correlation coefficients between CNCI and AP scores and their rankings for given sample universities within multi-year group (in 2014-2020)

| Schema | Document type | Sample size | r | p-value(r) | rho | p-value(rho) |
|---|---|---|---|---|---|---|
| ESI | Type-Filtered | Top100 | 0.940** | <0.001 | 0.948** | <0.001 |
| | | Top1000 | 0.869** | <0.001 | 0.889** | 0.000 |
| | | Top2000 | 0.802** | 0.000 | 0.853** | 0.000 |
| | Type-Unfiltered | Top100 | 0.222* | 0.027 | 0.096 | 0.343 |
| | | Top1000 | 0.493** | <0.001 | 0.511** | <0.001 |
| | | Top2000 | 0.582** | <0.001 | 0.630** | <0.001 |
| WoS | Type-Filtered | Top100 | 0.884** | <0.001 | 0.927** | <0.001 |
| | | Top1000 | 0.865** | <0.001 | 0.903** | 0.000 |
| | | Top2000 | 0.804** | 0.000 | 0.863** | 0.000 |
| | Type-Unfiltered | Top100 | 0.453** | <0.001 | 0.315* | 0.001 |
| | | Top1000 | 0.640** | <0.001 | 0.631** | <0.001 |
| | | Top2000 | 0.681** | <0.001 | 0.715** | 0.000 |

Note: The publication date for papers is set to 2014-2020.
Top 100-2000 refer as Table10.
**, Significant at the 0.01 significance level.
*, Significant at the 0.05 significance level.

Analysis of Tables 10 and 11 indicates that, when field classification schema, document type, and sample size are held constant, both the score and ranking correlations between CNCI and AP are lower in the multi-year group compared to the single-year group. Additionally, the data reveal a tendency regarding document type and schema: the ESI scheme yields higher correlations when the document type is restricted, whereas the WoS scheme performs better when no restrictions are applied. However, this trend diminishes with increasing sample size. For example, in both single-year and multi-year groups, when the sample size reaches 1,000 or 2,000 universities and the document type is restricted, the ranking correlation of CNCI and AP scores with the WoS scheme exceeds that of the ESI scheme.

From Table 10 and 11, we also can observe that whether the document types are filtered is the most significant factor influencing the correlation between the CNCI and AP scores and their rankings. For instance, as shown in Table 10, under the ESI scheme with document type filtering, correlation coefficients for scores and rankings were no less than 0.878 ($p = 0.000$) across all sample sizes. When document types were not filtered, however, the correlation strength dropped notably, with a maximum coefficient of only 0.739 ($p = 0.000$). This contrast confirms the critical role of document type filtering.

This difference can be more clearly observed in multi-year group. Within the ESI field classification scheme, when document type was filtered, the correlation coefficients between CNCI and AP scores were all greater than 0.8. In contrast, when unfiltered, the correlation fell to 0.222 (*p = 0.027*), and the ranking correlation became non-significant (rho = 0.096, *p = 0.343*). Interestingly, we also found that, regardless of which field classification scheme we selected or whether the publication year was continuous or single, the correlation decreased with sample size when document type was filtered, whereas the opposite trend occurred when no filtering was applied.

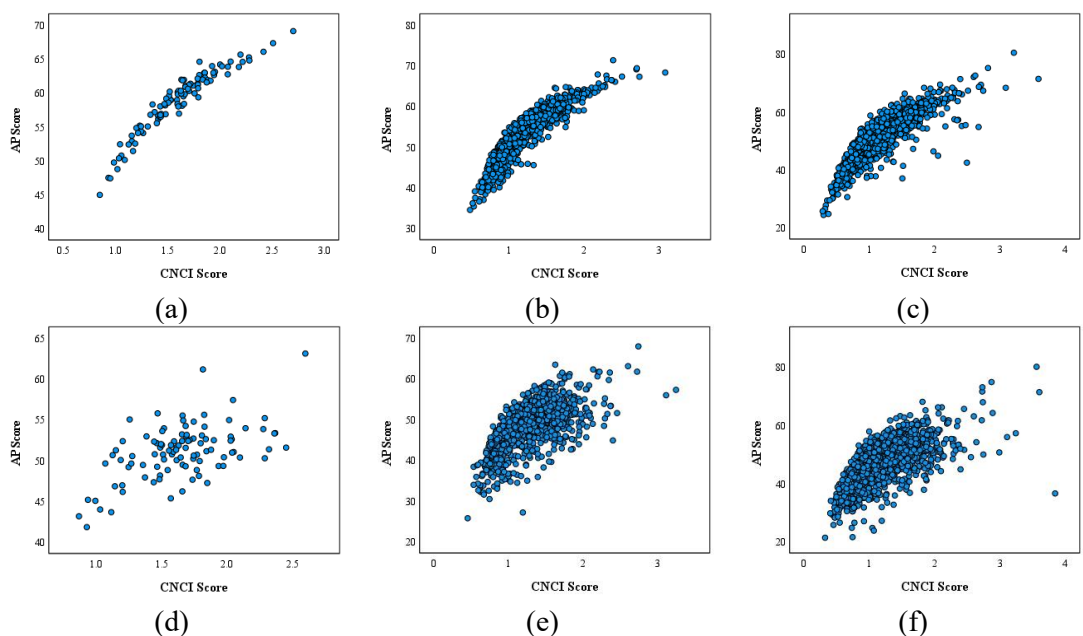

**Fig.6.** Correlation between the CNCI and AP scores for given sample universities with the ESI scheme in single-year group.
*Note:* (a)-(c) represent the correlation between CNCI and AP scores for the top 100, top 1,000, and top 2,000 universities with filtered document types; (d)-(f) represent the correlation for the given sample universities with unfiltered document types.

To furtherly explore correlation of CNCI and AP, we use scatter plots to visualize the distributions of their scores and rankings under filtered and unfiltered document conditions within the ESI scheme of single-year group. From fig. 6, we can observe that the scatter plot distributions for subgraph (a)- (c) (i.e. the type-filtered group) are relatively concentrated. The shape of these scatter plots is similar to the results of our previous experiment (see in Fig.3, d-f), showing a roughly diagonal distribution. This distribution offers an intuitive explanation for the high correlation reported in previous findings, which were based exclusively on the "article" document type. In contrast, the unfiltered condition (Fig.3, d–f) shows irregular distributions that grow denser with sample size. This more clearly illustrates that, after filtering the document type, there is a stronger correlation between CNCI and AP scores. In contrast, without filtering, the correlation between the two indicators is weaker, which is consistent with the conclusions drawn earlier.

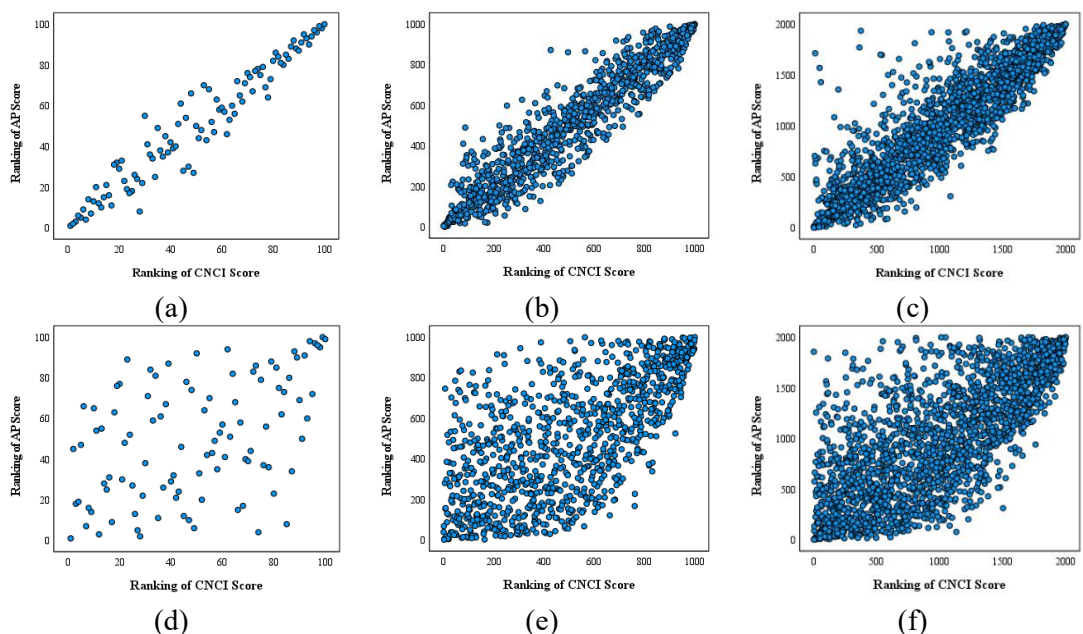

**Fig.7.** Correlation between the rankings of CNCI and AP scores for given sample universities with the ESI scheme in single-year group.
*Note:* (a)- (c) represent the correlation between the rankings of the CNCI and AP scores for top 100, top1,000 and top 2,000 universities with the filtered literature type; (d)- (f) represent the correlation for the given sample universities with unfiltered document types.

In Fig. 7, we can also find that, when the document type is filtered (see in Fig.7, a- c), the scatter plot distribution is tight and generally follows a straight-line pattern. In contrast, in Fig.7 (d)- (f), the scatter plot distribution is more dispersed, where the document type is not filtered.

Further analysis of Fig. 6 and 7 reveals how sample size influences distribution patterns. In Fig.6 and 7 (a) - (c), we can see that as the sample size increases, the number of outliers also increases. This trend explains our earlier finding of an inverse correlation between sample size and the CNCI-AP correlation when document types are filtered. Conversely, in subplots (d)-(f), the scatter plots grow denser with increasing sample size, which clarifies the positive correlation with sample size observed under unfiltered conditions in Tables 10 and 11. Additionally, we can see that the overall shape of the three subplots is similar. We believe that the increased correlation between the indicators as the university sample size grows is a natural result of the data accumulation effect.

Our analysis demonstrates that document type is the key factor affecting the correlation between nonlinear and linear normalization scores, outweighing the influence of publication year and field classification system. Filtering document types reduces data heterogeneity, allowing normalization methods to function more effectively and thereby strengthening score correlations. Additionally, the impact of sample size on the correlation between indicators depends on the homogeneity of the data. These findings highlight that document type filtering is crucial for ensuring stable and consistent citation-based evaluations, providing key insights for selecting and interpreting normalization methods.

## 5 Conclusion

This study empirically investigates the errors introduced by summing and averaging nonlinearly field-normalized citation counts, a common but mathematically problematic practice in evaluative bibliometrics. Through single-field and cross-field experiments that applied both linear (e.g., Mean-based, Median-based, Z-score) and nonlinear (e.g., Percentile rank, NLCS, Logarithmic z-score) field normalization methods to real citation data, we obtained several key conclusions.

Our experiments confirm that nonlinear normalization methods violate the equal-interval property of raw citation data. As theorized by Wang (2024), this violation introduces systematic errors when such data are summed or averaged, in contrast to linear methods. However, our findings further reveal that the practical impact of these errors varies significantly across practical contexts.

In the single-field experiment, the errors are relatively small in practice, as indicated by the high correlation between normalized scores and raw citations. Nevertheless, it is visually evident that nonlinearly normalized data exhibits a more pronounced deviation from raw citation counts compared to linearly normalized data. In the cross-field experiment, which more closely simulates real evaluation, the errors are even smaller. This can be partly explained by the poor performance of linear reference methods like the Z-score on skewed data, which narrows the practical error between linear and nonlinear approaches and maintains a high correlation between their scores.

We further investigated the discrepancy between our findings and those reported by Wang (2024), who observed a low correlation between linear and nonlinear indicators. Our analysis identified document type filtering as a decisive factor: restricting the selection to "articles" significantly strengthens the correlation between linear and nonlinear indicators, whereas including all document types increases error and weakens their correlation. This highlights the crucial role of data preparation in bibliometric studies.

Our findings confirm the theoretical drawback of nonlinear normalization methods and demonstrate their impact in cross-field assessments. However, their inherent violation of additivity cannot be overlooked. In summary, evaluations that aggregate citation data must choose mathematically appropriate normalization methods for summing or averaging. If nonlinear indicators are applied, their flaws must be clear stated. The strong influence of data preparation choices (e.g., document type inclusion) must be accounted for, as they are essential for obtaining reliable and meaningful findings.

While the study's findings are noteworthy, this study has several limitations that suggest potential directions for future research.

First, the analysis was conducted primarily at the university level. Future work could explore other levels (e.g., national, journal, research team, or individual) to assess the broader applicability of our findings.

Second, our study focused on internal mathematical consistency, using raw or linear normalized citation counts as the benchmark, but did not validate the results against an external quality measure like peer review. Future studies, as suggested by Bornmann and Haunschild (2024), could examine whether the errors we identified lead to assessments that diverge from expert judgment.

Secondly, our study primarily focused on the internal consistency of mathematical properties, using raw citation data or linear normalized citation data as benchmarks. We did not correlate our results with an external quality benchmark, such as peer review outcomes. Future research, as suggested by Bornmann and Haunschild (2024), could be to examine whether the errors we identified actually lead to different or less accurate assessments of research quality compared to expert judgment.

Thirdly, our analysis of influencing factors was not exhaustive. Future work could systematically investigate the impact of other variables, such as the length of the citation window, different field classification systems beyond ESI and WoS, or the effect of highly cited papers on the aggregation results.

Finally, future research could attempt to explore more practical methods to guide evaluators. This would help minimize the impact of evaluation errors, ensuring that such errors do not unfairly affect the outcomes of funding, ranking, or promotion decisions.

## Acknowledgment

My sincere thanks go to my supervisor, Dr. Xing Wang, for his expert guidance and constructive feedback, both of which greatly improved the quality of this work.

# Supplementary materials

Supplementary material associated with this article can be found, in the online version, at
…

# Appendix A: Other figures in single-field experiment

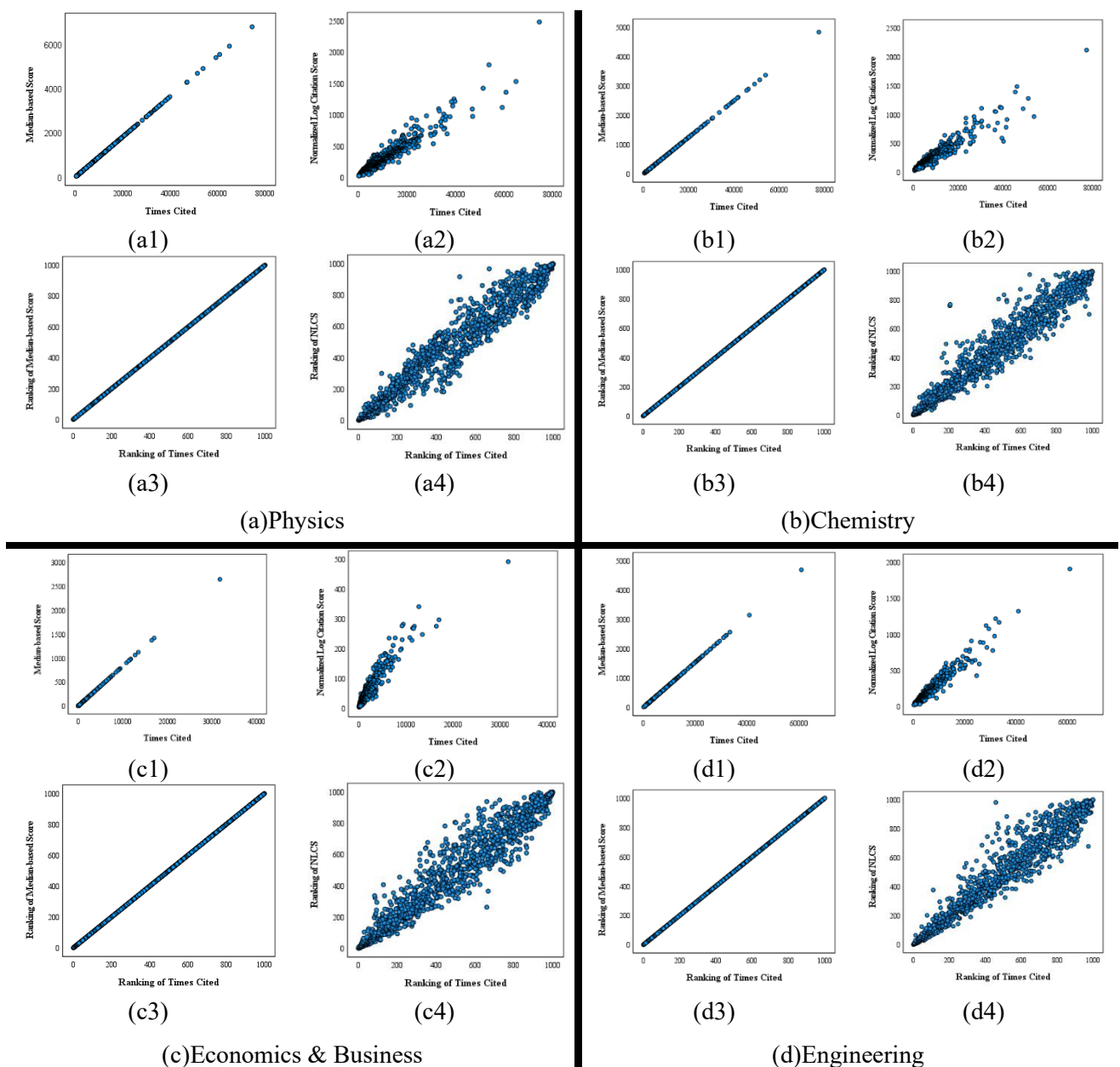

**Fig.8.** Correlation between the **summarized Median-based and NLCS score** against raw citation counts and correlation between the rankings of these indicators for sample universities in different fields.

Note: 1. (a1)- (a2), (b1)- (b2), (c1)- (c2), (d1)- (d2) represent of the correlation of the Median-based and NLCS *score against raw citation counts* for sample universities in Physics, Chemistry, Economics & Business and Engineering, respectively;

2. (a3)- (a4), (b3)- (b4), (c3)- (c4), (d3)- (d4) represent the correlation of the Median-based and NLCS score' *rankings versus raw citation counts' rankings* for sample universities in Physics, Chemistry, Economics & Business, and Engineering, respectively.

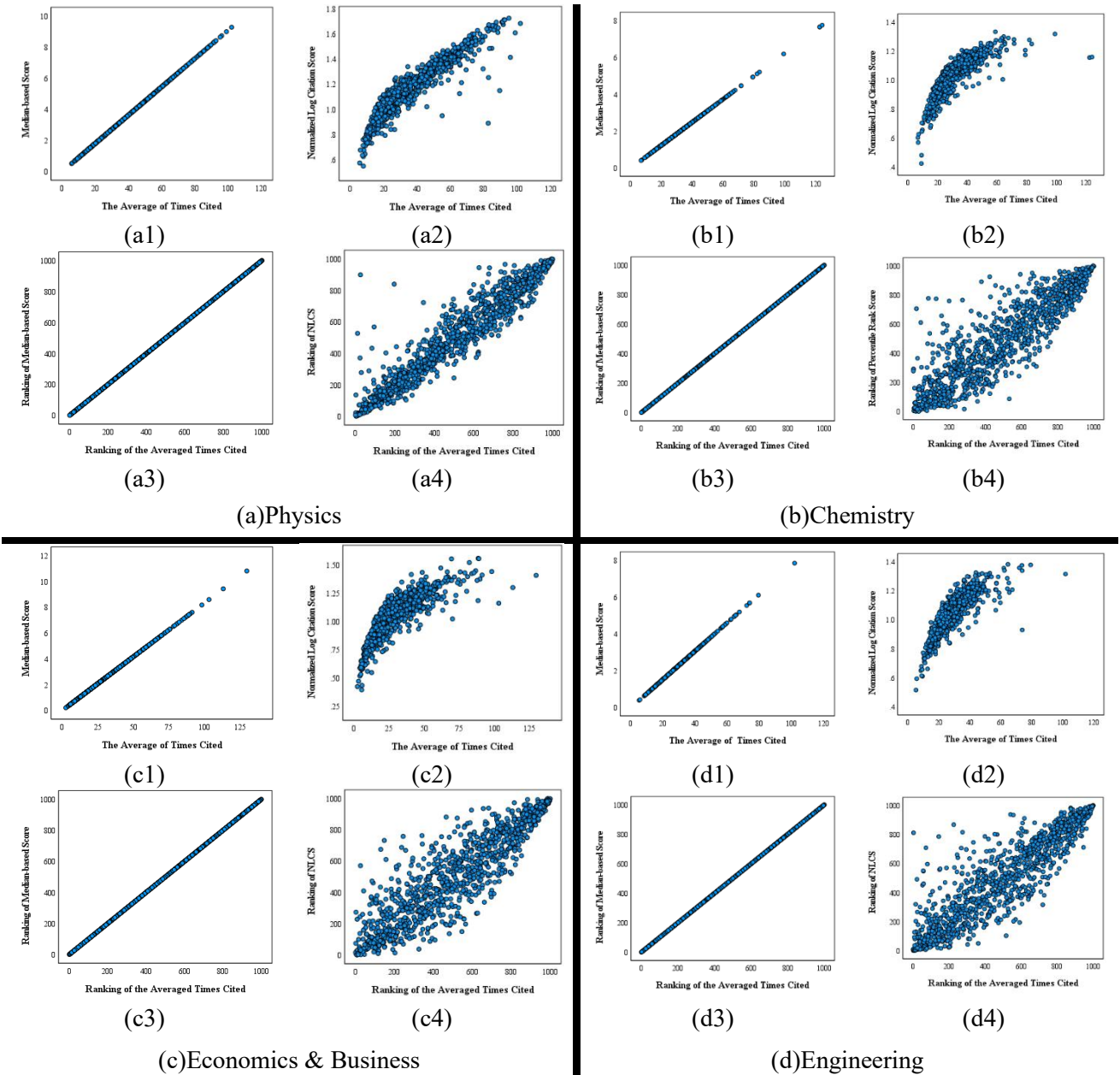

**Fig.9.** Correlation between the **averaged Median-based and NLCS score** against the average of raw citation counts and correlation between the rankings of these indicators for sample universities in different fields. Note: (a1)- (d4) represent correlations consistent with Fig.8.

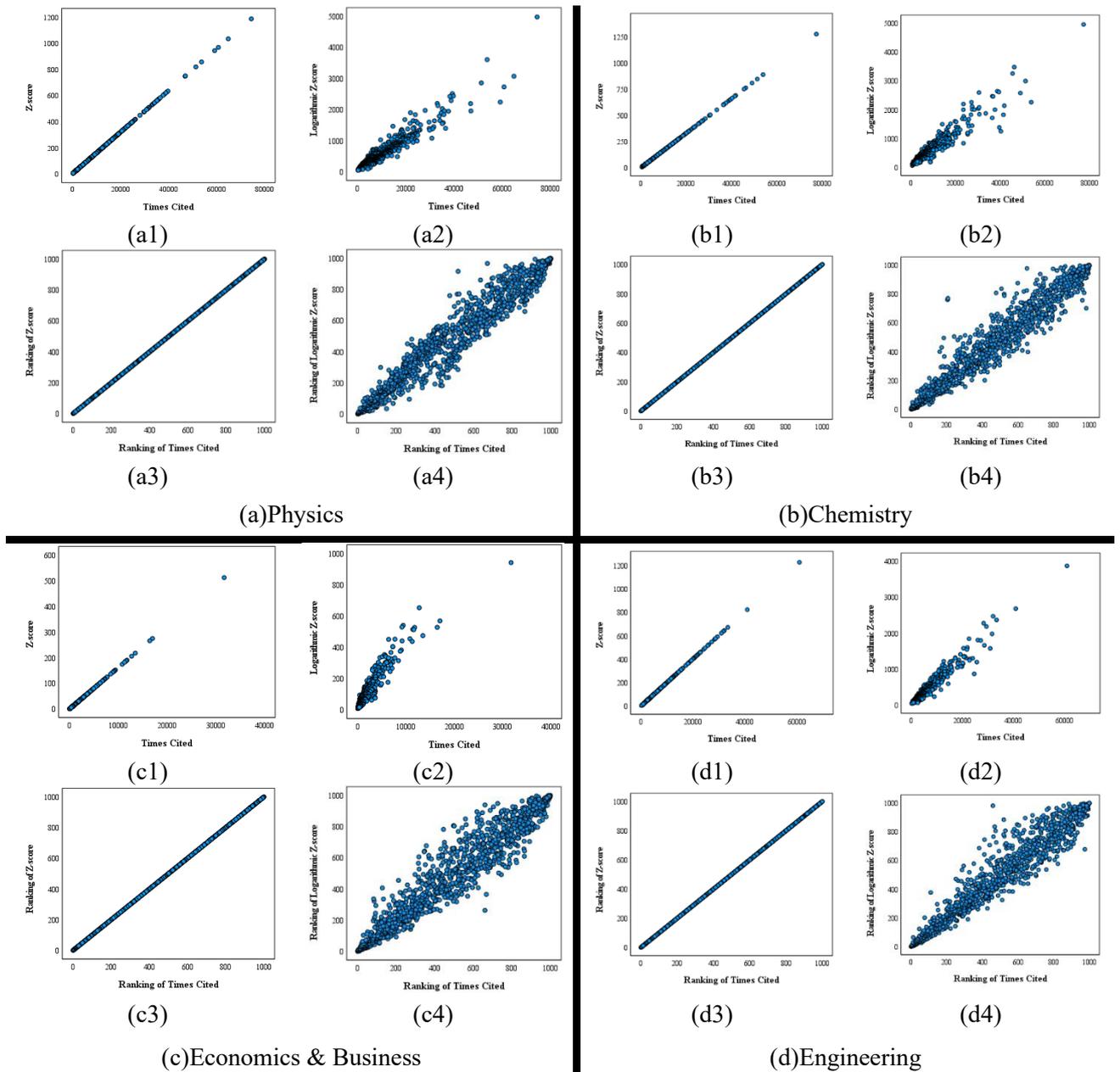

**Fig.10.** Correlation between the **summarized Z-score and Log z-score** against the raw citation counts and correlation between the rankings of these indicators for sample universities in different fields.

Note: 1. (a1)- (a2), (b1)- (b2), (c1)- (c2), (d1)- (d2) represent of the correlation of *the Z score and Log z score against raw citation counts* for sample universities in Physics, Chemistry, Economics & Business and Engineering, respectively;

2. (a3)- (a4), (b3)- (b4), (c3)- (c4), (d3)- (d4) represent the correlation of *the Z score and Log z score' rankings versus raw citation counts' rankings* for sample universities in Physics, Chemistry, Economics & Business, and Engineering, respectively.

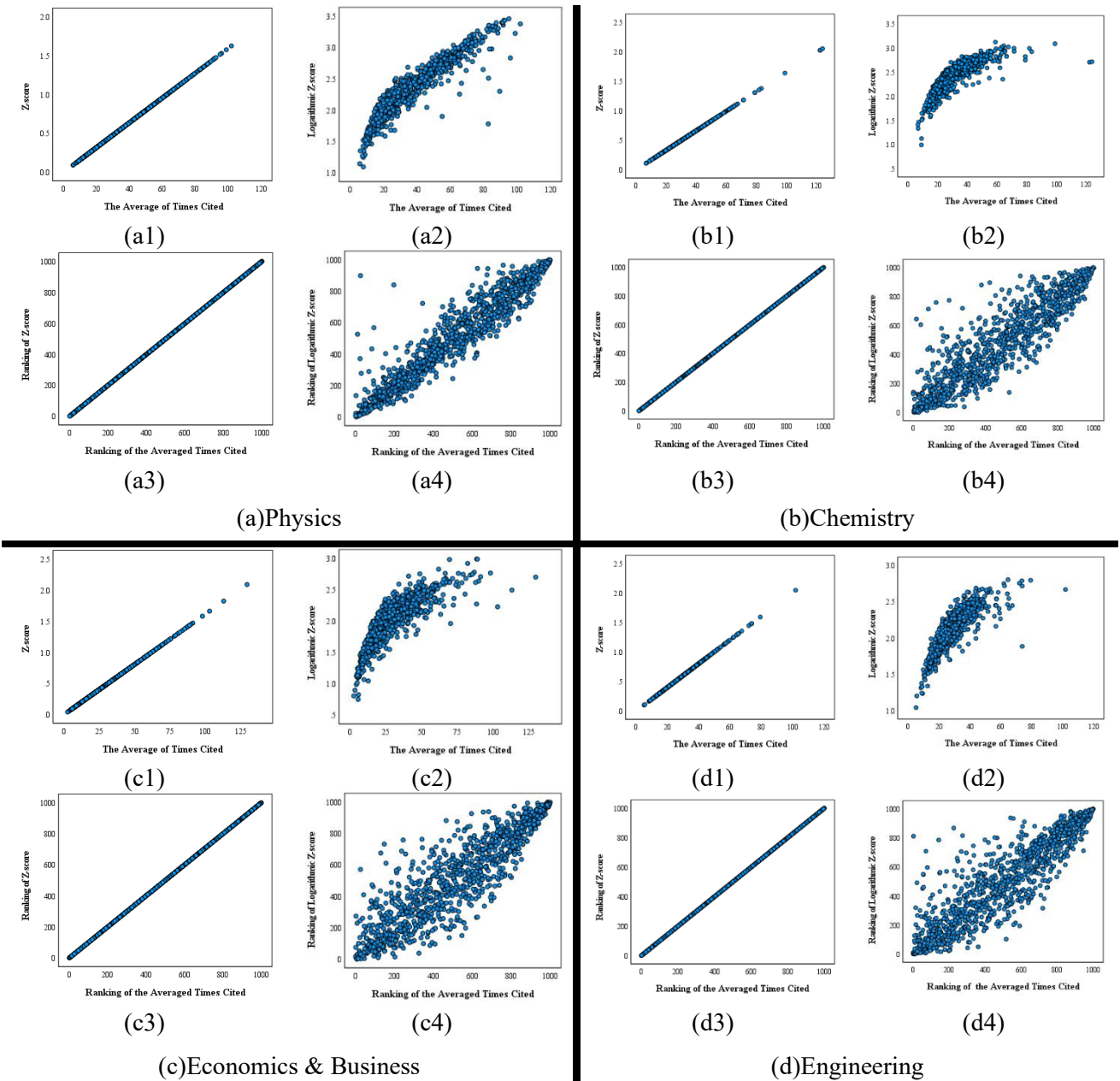

**Fig.11.** Correlation between the **averaged Z-score and Log z-score** against the average of raw citation counts and correlation between the rankings of these indicators for sample universities in different fields.
Note: (a1)- (d4) represent correlations consistent with Fig.10.

**Table 12**
Ranking changes of the Top 20 universities with the most articles in Physics after **averaging** the normalization citation counts

| No. | University Name | Article Number | TC | TC (ave) | Rank0 | Median score | Rank1 | Rank C1 | Z score | Rank2 | Rank C2 |
|---|---|---|---|---|---|---|---|---|---|---|---|
| 0001 | Universite Paris Saclay | 2055 | 74509 | 36.257 | 468 | 3.296 | 468 | 0 | 0.578 | 468 | 0 |
| 0002 | University of Tokyo | 1513 | 53795 | 35.555 | 480 | 3.232 | 480 | 0 | 0.567 | 480 | 0 |
| 0003 | Sorbonne Universite | 1122 | 51408 | 45.818 | 346 | 4.165 | 346 | 0 | 0.73 | 346 | 0 |
| 0004 | University of Science & Technology of China, CAS | 1115 | 38372 | 34.414 | 502 | 3.129 | 502 | 0 | 0.548 | 502 | 0 |
| 0005 | Massachusetts Institute of Technology (MIT) | 1113 | 64914 | 58.323 | 165 | 5.302 | 165 | 0 | 0.929 | 165 | 0 |
| 0006 | Tsinghua University | 1111 | 39388 | 35.453 | 482 | 3.223 | 482 | 0 | 0.565 | 482 | 0 |
| 0007 | Peking University | 1069 | 39782 | 37.214 | 453 | 3.383 | 453 | 0 | 0.593 | 453 | 0 |
| 0008 | University of California Berkeley | 1005 | 60793 | 60.491 | 143 | 5.499 | 143 | 0 | 0.964 | 143 | 0 |
| 0009 | Lomonosov Moscow State University | 999 | 33418 | 33.451 | 513 | 3.041 | 513 | 0 | 0.533 | 513 | 0 |
| 0010 | Universite Grenoble Alpes (UGA) | 929 | 33126 | 35.658 | 478 | 3.242 | 478 | 0 | 0.568 | 478 | 0 |
| 0011 | Tohoku University | 906 | 22014 | 24.298 | 681 | 2.209 | 681 | 0 | 0.387 | 681 | 0 |
| 0012 | University of Chinese Academy of Sciences, CAS | 878 | 19736 | 22.478 | 723 | 2.043 | 723 | 0 | 0.358 | 723 | 0 |
| 0013 | University of Cambridge | 870 | 38904 | 44.717 | 360 | 4.065 | 360 | 0 | 0.713 | 360 | 0 |
| 0014 | Kyoto University | 861 | 28123 | 32.663 | 523 | 2.969 | 523 | 0 | 0.52 | 523 | 0 |
| 0015 | Osaka University | 835 | 24028 | 28.776 | 590 | 2.616 | 590 | 0 | 0.459 | 590 | 0 |
| 0016 | Stanford University | 809 | 59239 | 73.225 | 49 | 6.657 | 49 | 0 | 1.167 | 49 | 0 |
| 0017 | University of Oxford | 803 | 46940 | 58.456 | 161 | 5.314 | 161 | 0 | 0.931 | 161 | 0 |
| 0018 | Universite Paris Cite | 778 | 35222 | 45.272 | 355 | 4.116 | 355 | 0 | 0.721 | 355 | 0 |
| 0019 | University of Maryland College Park | 750 | 36078 | 48.104 | 304 | 4.373 | 304 | 0 | 0.767 | 304 | 0 |
| 0020 | Huazhong University of Science & Technology | 748 | 14774 | 19.751 | 776 | 1.796 | 776 | 0 | 0.315 | 776 | 0 |

Note: *No. 0001-0020* represent the top 20 universities with the highest publication outputs in Physics; *Rank0-2* respectively refer to the rankings of average of raw citation counts (TC ave), median-based score and Z score; *Rank C1-2* respectively refer to the changes in rankings of median-based score and Z score against the average of raw citation counts' rankings (Rank0).

**Table 13**
Ranking changes of the Top 20 universities with the most articles in Physics after **summing** the normalization citation counts

| No. | University Name | Article Number | TC | Rank0 | Median score | Rank1 | Rank C1 | Z score | Rank2 | Rank C2 |
|---|---|---|---|---|---|---|---|---|---|---|
| 0001 | Universite Paris Saclay | 2055 | 74509 | 1 | 6773.545 | 1 | 0 | 1187.292 | 1 | 0 |
| 0002 | University of Tokyo | 1513 | 53795 | 5 | 4890.455 | 5 | 0 | 857.217 | 5 | 0 |
| 0003 | Sorbonne Universite | 1122 | 51408 | 6 | 4673.455 | 6 | 0 | 819.181 | 6 | 0 |
| 0004 | University of Science & Technology of China, CAS | 1115 | 38372 | 12 | 3488.364 | 12 | 0 | 611.453 | 12 | 0 |
| 0005 | Massachusetts Institute of Technology (MIT) | 1113 | 64914 | 2 | 5901.273 | 2 | 0 | 1034.397 | 2 | 0 |
| 0006 | Tsinghua University | 1111 | 39388 | 10 | 3580.727 | 10 | 0 | 627.643 | 10 | 0 |
| 0007 | Peking University | 1069 | 39782 | 9 | 3616.545 | 9 | 0 | 633.922 | 9 | 0 |
| 0008 | University of California Berkeley | 1005 | 60793 | 3 | 5526.636 | 3 | 0 | 968.729 | 3 | 0 |
| 0009 | Lomonosov Moscow State University | 999 | 33418 | 23 | 3038 | 23 | 0 | 532.512 | 23 | 0 |
| 0010 | Universite Grenoble Alpes (UGA) | 929 | 33126 | 24 | 3011.455 | 24 | 0 | 527.859 | 24 | 0 |
| 0011 | Tohoku University | 906 | 22014 | 62 | 2001.273 | 62 | 0 | 350.791 | 62 | 0 |
| 0012 | University of Chinese Academy of Sciences, CAS | 878 | 19736 | 77 | 1794.182 | 77 | 0 | 314.491 | 77 | 0 |
| 0013 | University of Cambridge | 870 | 38904 | 11 | 3536.727 | 11 | 0 | 619.931 | 11 | 0 |
| 0014 | Kyoto University | 861 | 28123 | 33 | 2556.636 | 33 | 0 | 448.137 | 33 | 0 |
| 0015 | Osaka University | 835 | 24028 | 48 | 2184.364 | 48 | 0 | 382.883 | 48 | 0 |
| 0016 | Stanford University | 809 | 59239 | 4 | 5385.364 | 4 | 0 | 943.967 | 4 | 0 |
| 0017 | University of Oxford | 803 | 46940 | 8 | 4267.273 | 8 | 0 | 747.983 | 8 | 0 |
| 0018 | Universite Paris Cite | 778 | 35222 | 19 | 3202 | 19 | 0 | 561.259 | 19 | 0 |
| 0019 | University of Maryland College Park | 750 | 36078 | 15 | 3279.818 | 15 | 0 | 574.899 | 15 | 0 |
| 0020 | Huazhong University of Science & Technology | 748 | 14774 | 150 | 1343.091 | 150 | 0 | 235.422 | 150 | 0 |

Note: *No. 0001-0020* as described in Table 12. *Rank0-2* respectively refer to the rankings of raw citation counts (TC), median-based score and Z score; *Rank C1-2* respectively refer to the changes in rankings of median-based score and Z score against the raw citation counts' rankings (Rank0).

# Appendix B: Other figures in cross-field experiment

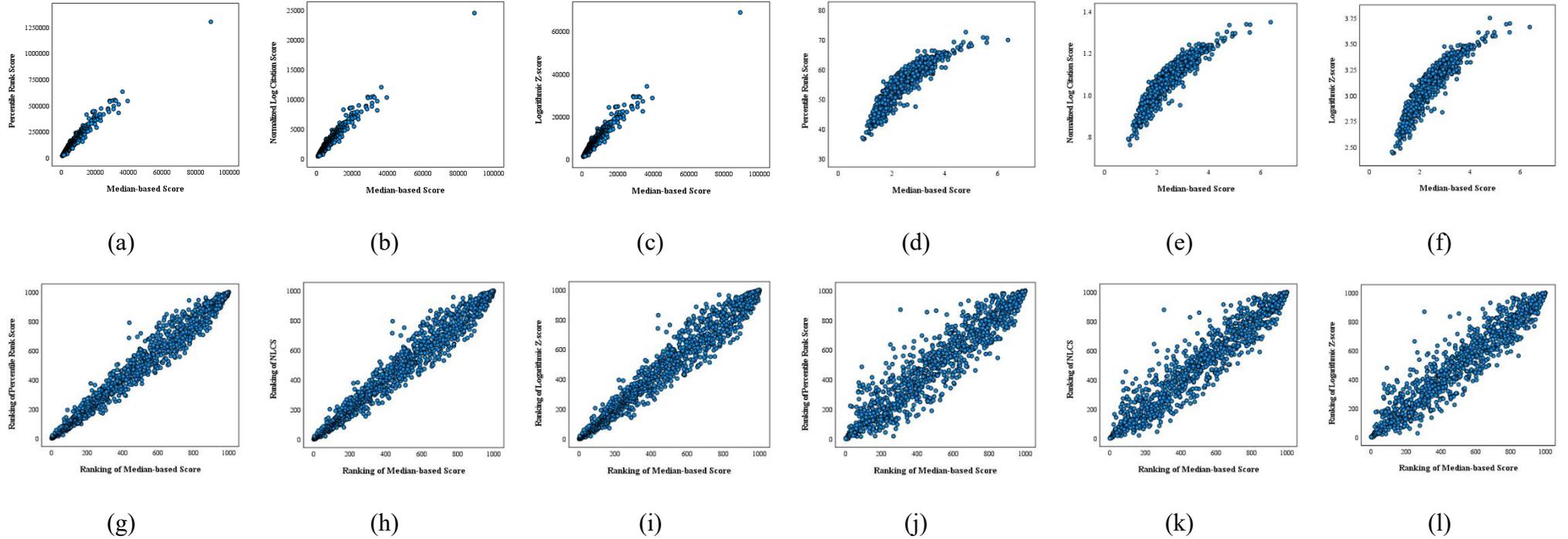

**Fig.12.** Correlation of **Median-based score** against nonlinear field normalization score and the correlation of their rankings for all sample universities

Note: 1. (a)-(c) shows the correlation of the Median-based score against nonlinear field normalization *score by summing* ($AP_1$); (d)-(f) shows the correlation of the Median-based score against nonlinear field normalization *score by averaging* ($AP_2$).
2. (g)-(i) shows the *ranking distribution* among the Median-based score and nonlinear field normalization indicators *by summing* ($AP_1$); (j)-(l) shows the *ranking distribution* among the Median-based score and nonlinear field normalization indicators *by averaging* ($AP_2$).

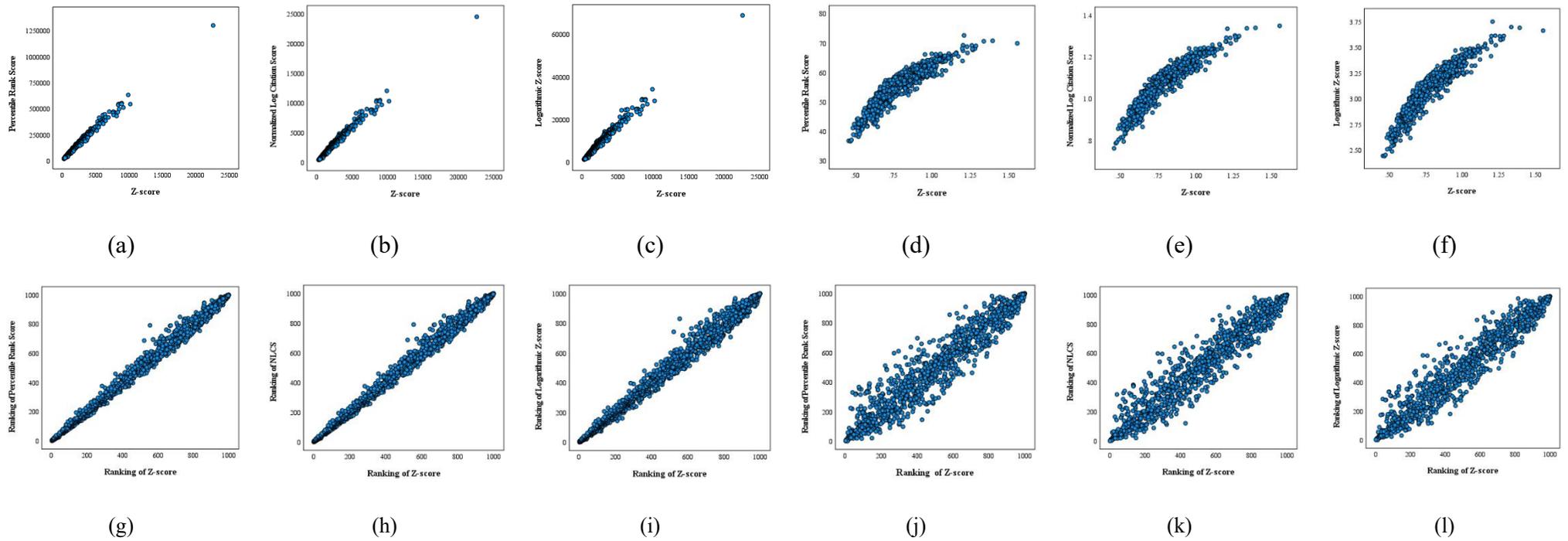

**Fig.13.** Correlation of **Z-score** against nonlinear field normalization score and the correlation of their rankings for all sample universities

Note: 1. (a)-(c) shows the correlation of the Z-score against nonlinear field normalization *score by summing* (AP$_1$); (d)-(f) shows the correlation of the Z- score against nonlinear field normalization *score by averaging* (AP$_2$).
2. (g)-(i) shows the *ranking distribution* among the Z-score and nonlinear field normalization indicators *by summing* (AP$_1$); (j)-(l) shows the *ranking distribution* among the Z-score and nonlinear field normalization indicators *by averaging* (AP$_2$).

**Table 14**
Pearson correlations coefficient between **summarized** linear and nonlinear field normalized scores and Spearman correlation coefficient between the rankings of these indicators for Top 100 universities.

| Linear | Nonlinear | Sample size | r | P-value(r) | rho | P-value(rho) |
|---|---|---|---|---|---|---|
| Mean-based score | Percentile Rank score | 100 | .962** | <0.001 | .923** | <0.001 |
| | NLCS | 100 | .956** | <0.001 | .908** | <0.001 |
| | Log z-score | 100 | .948** | <0.001 | .889** | <0.001 |
| Median-based score | Percentile Rank score | 100 | .961** | <0.001 | .920** | <0.001 |
| | NLCS | 100 | .955** | <0.001 | .905** | <0.001 |
| | Log z-score | 100 | .946** | <0.001 | .886** | <0.001 |
| Z-score | Percentile Rank score | 100 | .986** | <0.001 | .977** | <0.001 |
| | NLCS | 100 | .983** | <0.001 | .967** | <0.001 |
| | Log z-score | 100 | .977** | <0.001 | .954** | <0.001 |

Note: Linear and Nonlinear as defined in Table 7.
**. Significant at the 1% significance level.

**Table 15**
Pearson correlation coefficient between **averaged** linear and nonlinear field normalized scores and Spearman correlation coefficient between the rankings of these indicators for Top 100 universities.

| Linear | Nonlinear | Sample size | r | P-value(r) | rho | P-value(rho) |
|---|---|---|---|---|---|---|
| Mean-based score | Percentile Rank score | 100 | .946** | <0.001 | .960** | <0.001 |
| | NLCS | 100 | .955** | <0.001 | .965** | <0.001 |
| | Log z-score | 100 | .959** | <0.001 | .967** | <0.001 |
| Median-based score | Percentile Rank score | 100 | .939** | <0.001 | .952** | <0.001 |
| | NLCS | 100 | .951** | <0.001 | .960** | <0.001 |
| | Log z-score | 100 | .952** | <0.001 | .960** | <0.001 |
| Z-score | Percentile Rank score | 100 | .962** | <0.001 | .969** | <0.001 |
| | NLCS | 100 | .970** | <0.001 | .975** | <0.001 |
| | Log z-score | 100 | .971** | <0.001 | .974** | <0.001 |

Note: Linear and Nonlinear as defined in Table 7.
**. Significant at the 1% significance level.